\documentclass[final,5p,times,twocolumn,authoryear]{elsarticle}

\myfooter[L]{Preprint version}

\usepackage{amssymb}

\usepackage{lineno}
\usepackage{caption, subcaption}
\usepackage{color, graphicx}
\usepackage{float}
\usepackage{amssymb}
\usepackage{amsmath}
\usepackage{mathabx}
\usepackage{booktabs}
\graphicspath{{figures/}}
\usepackage{epstopdf}
\usepackage{color,soul}
\usepackage{siunitx}
\usepackage[shortlabels]{enumitem}
\usepackage{multirow}

\usepackage{siunitx}
\usepackage{soul,color}
\soulregister\cite7
\soulregister\ref7
\soulregister\pageref7
\soulregister\citep7
\usepackage{url}  
\usepackage[shortlabels]{enumitem}
\usepackage{adjustbox}
\usepackage{siunitx}

\usepackage{titlesec}

\titlespacing*{\subsection}
  {0pt}    
  {0.5\baselineskip} 
  {0.25\baselineskip} 

\usepackage{setspace}

\begin{document}

\begin{frontmatter}



\title{From large-eddy simulations to deep learning: A U-net model for fast urban canopy flow predictions}





\author[y2e2]{
Themistoklis Vargiemezis\corref{cor1}}
\emailauthor{tvarg@stanford.edu} {Themistoklis Vargiemezis}

\author[y2e2]{Catherine Gorl\'e}

\address[y2e2]{Stanford University, Y2E2 Building, 473 Via Ortega, Stanford, CA, 94305}

\begin{abstract}
Accurate prediction of wind flow fields in urban canopies is crucial for ensuring pedestrian comfort, safety, and sustainable urban design. Traditional methods using wind tunnels and Computational Fluid Dynamics, such as Large-Eddy Simulations (LES), are limited by high costs, computational demands, and time requirements. This study presents a deep neural network (DNN) approach for fast and accurate predictions of urban wind flow fields, reducing computation time from $\mathcal{O}10$ hours on 32 CPUs for one LES evaluation to $\mathcal{O}1$ seconds on a single GPU using the DNN model. We employ a U-Net architecture trained on LES data including 252 synthetic urban configurations at seven wind directions ($0^{\circ}$ to $90^{\circ}$ in $15^{\circ}$ increments). The model predicts two key quantities of interest: mean velocity magnitude and streamwise turbulence intensity, at multiple heights within the urban canopy. The U-net uses 2D building representations augmented with signed distance functions and their gradients as inputs, forming a $256\times256\times9$ tensor. In addition, a Spatial Attention Module is used for feature transfer through skip connections. The loss function combines the root-mean-square error of predictions, their gradient magnitudes, and L2 regularization. Model evaluation on 50 test cases demonstrates high accuracy with an overall mean relative error of 9.3\% for velocity magnitude and 5.2\% for turbulence intensity. This research shows the potential of deep learning approaches to provide fast, accurate urban wind assessments essential for creating comfortable and safe urban environments. Code is available at \url{https://github.com/tvarg/Urban-FlowUnet.git}

\end{abstract}



\end{frontmatter}



\section{Introduction}

Pedestrian wind comfort is a critical aspect of urban design, influencing both the safety and livability of city environments. High wind speeds can lead to discomfort, hinder mobility, and even pose serious hazards to pedestrians. Therefore, the ability to accurately and quickly determine flow fields in urban areas is essential for ensuring pedestrian wind comfort. Wind-induced interference effects can significantly change not only the flow field but also the pressure loads on a building, and site-specific studies are required to assess safety and wind resiliency in urban areas. 

In the quest for enhancing urban wind comfort and safety, the fast and accurate prediction of flow fields plays a pivotal role. Traditionally, this analysis has relied heavily on the use of wind tunnels \citep{blocken2016pedestrian, tieleman2003wind, zhang2008wind, plate2001wind, phillips2019will, vargiemezis2023experimental}, which, despite their effectiveness, come with a set of substantial limitations. These limitations include their general unavailability due to high demand, the high operational costs associated with their use, and the logistical challenge of testing every possible configuration to simulate real-world scenarios. As urban areas continue to grow and evolve, finding more accessible and scalable solutions has become increasingly urgent.

Recently, Computational Fluid Dynamics (CFD) has gained attention as a promising tool for studying wind effects in urban environments, with Reynolds-Averaged Navier-Stokes (RANS) modeling and Large Eddy Simulation (LES) being two widely used approaches. RANS remains the most commonly applied method due to its relatively low computational cost, making it suitable for wind engineering applications \citep{van2010coupled, yoshie2007cooperative, montazeri2013cfd, janssen2013pedestrian, kang2020computational, bacs2024city}. However, it relies on turbulence models that may introduce inaccuracies in complex, highly unsteady flows due to modeling all turbulent scales using phenomenological models. In contrast, LES offers a more detailed analysis by directly resolving large-scale turbulent structures, leading to higher accuracy in cases where transient effects are of great importance. LES has been successfully used to predict pedestrian wind flows and pollutant dispersion in urban areas \citep{moonen2013performance, salim2011numerical, adamek2017pedestrian}, to model wind-driven ventilation in buildings \citep{hwang2022large, hwang2023large, zhang2021cfd}, and to investigate pressure loads on realistic low-rise buildings \citep{vargiemezis2024isolated, vargiemezis2024urban}. However, LES is computationally intensive and requires significant resources and time, which limits its widespread use in wind engineering practices \citep{blocken2015computational, potsis2023computational}.

In response to these challenges, data-driven approaches are gaining traction within the field of CFD. This method employs supervised learning algorithms based on deep neural networks (DNNs) for fast flow field predictions. By training these networks on large datasets generated from simulations, this approach can predict wind patterns and behaviors with high speed and accuracy. The fact that LES can provide detailed and accurate datasets of the flow field in such cases enables the use of DNN models with these data. The integration of deep learning into flow field analysis makes it possible to conduct assessments that are not only fast but also cost-effective. DNN models have been used successfully for fast approximation of turbulent shear flows \citep{srinivasan2019predictions}, to study airfoil aerodynamics \citep{saetta2024uncertainty, thuerey2020deep}, and flows over automobile vehicles \citep{chen2022flowdnn}.

In this study, we use a DNN model for fast and accurate predictions of flow fields in urban areas. The model is trained using a dataset created from LES, ensuring that it captures the complexities of urban wind flows. To build this dataset, we generate city models scaled down to a 1:100 ratio, where each full-scale model represents a $600m \times 600m$ urban area. These models consist of 70 randomly placed buildings, with a minimum spacing of 10 meters between any two structures. The dataset includes 252 different city geometries, at different wind directions, ranging from $0^\circ$ to $90^\circ$ in $15^\circ$ increments. The key contribution of this paper is the use of a DNN model for predictions that are less resource-intensive and provide urban planners and engineers with a powerful tool for predicting flow fields in urban areas,  which can potentially be used to optimize wind comfort and safety in city environments.

\section{Case study}
\subsection{Urban area geometries}
\begin{figure}[htbp]
    \centering
    \includegraphics[width=\columnwidth]{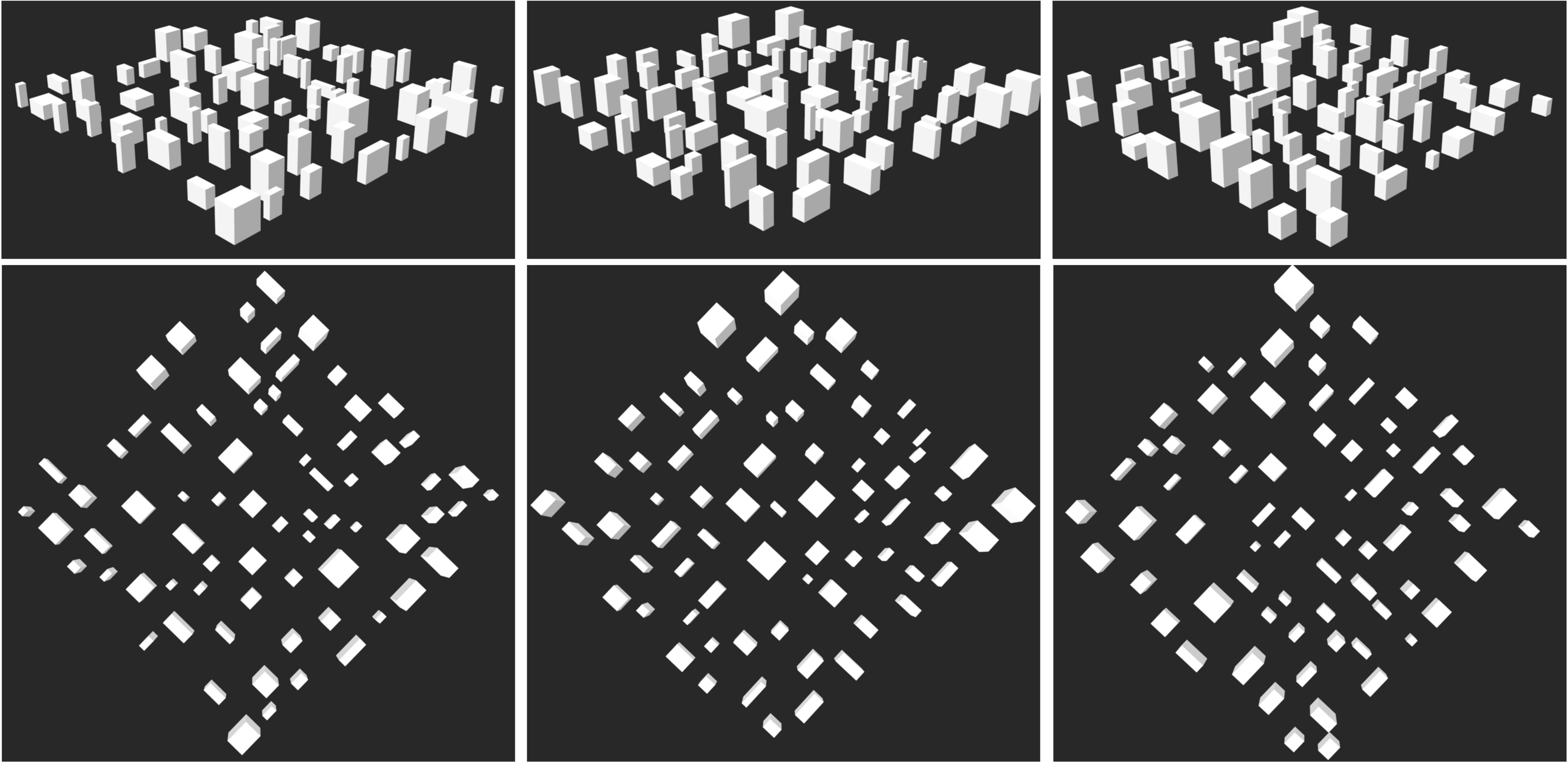}
    \caption{Examples of 3D urban models (top) and their top views (bottom)}
    \label{fig:city_models}
\end{figure}

In this study, we consider city models that are scaled down to 1:100 ratio. In full scale, each model represents a city spanning $600m \times 600m$. These models are populated with 70 randomly placed buildings within the specified area, with a minimum spacing of 10 meters between any two buildings. The dimensions of each building are randomly assigned, following a uniform distribution ranging from 10 to 50 meters ($L \sim \text{Uniform}(10, 50)$). Figure \ref{fig:city_models} shows three different configurations from a 3D perspective and their top view.

The database includes 252 different city geometries at 7 different wind directions ranging from $0^\circ$ to $90^\circ$, every $15^\circ$.  This approach divides the models into sets of 36 for each specified wind direction, ensuring that no city model is duplicated between different wind assessments. This diversity of configurations increases the robustness and accuracy of the model when predicting new, untested urban layouts.
\subsection{Quantities of interest}
We are interested in the prediction of flowfields within urban areas, which is especially important for assessing wind comfort in city environments. In this study, we focus on predicting two key quantities of interest (QoIs): 1) The mean velocity magnitude ${U}_{mag}$, and the streamwise turbulence intensity $I_u$. The turbulence intensity is given by the ratio of the root-mean-square wind speed and the mean velocity magnitude:
    \begin{equation}
    \label{eq:turb_int}
    I_u = \frac{\sigma_u}{{U}_{mag}}
    \end{equation}

These quantities are fundamental for pedestrian wind comfort assessment. Wind comfort criteria typically exist as a threshold value for the effective wind and a maximum allowed exceedence probability of this threshold. The effective wind, $U_e$,  which determines pedestrian comfort levels, is calculated using the two aforementioned QoIs \citep{bottema1999towards}:
    \begin{equation}
    \label{eq:comf_crit}
    U_e = {U}_{mag} + k \ \sigma_u \geq threshold,
    \end{equation}
where $k$ is the peak factor that takes values between 0 and 3.5 \citep{blocken2016pedestrian}. More detailed reviews on comfort criteria can be found in \citep{bottema2000method, janssen2013pedestrian}.

By accurately predicting ${U}_{mag}$ and $I_u$, researchers can calculate $U_e$ to asses pedestrian wind comfort levels and identity regions that may pose risks to pedestrians. While this important application lies beyond the scope of this paper, it represents a compelling avenue for future research. In this paper, our focus is on developing a DNN model to accurately predict these two fundamental QoIs.

\subsection{LES setup}
To perform the LES simulations and obtain the dataset for this study, we employ the CharLES code, a low-Mach, isentropic solver developed by \cite{charles}. CharLES implements a finite volume approach with an automated body-fitted meshing technique based on 3D-clipped Voronoi diagrams, and uses numerical schemes with second-order accuracy in space and time to solve the governing equations \citep{ambo2020aerodynamic}. In addition, the Vreman turbulence model is used for the unresolved part of the stress tensor \citep{vreman2004eddy}.  CharLES has been successfully used to study wind-induced pressure loads on realistic low-rise buildings in urban areas \citep{vargiemezis2024urban, vargiemezis2024isolated}, model natural ventilation in buildings within complex urban environments \citep{hwang2022large, hwang2023large}, and assess wind loading on high-rise buildings \citep{ciarlatani2023investigation}. This extensive validation across diverse wind engineering applications shows the reliability and versatility of CharLES for advanced wind flow simulations and obtaining datasets for training DNN models.
    \begin{figure}[htpb]
    \centering
    \includegraphics[width=1\columnwidth]{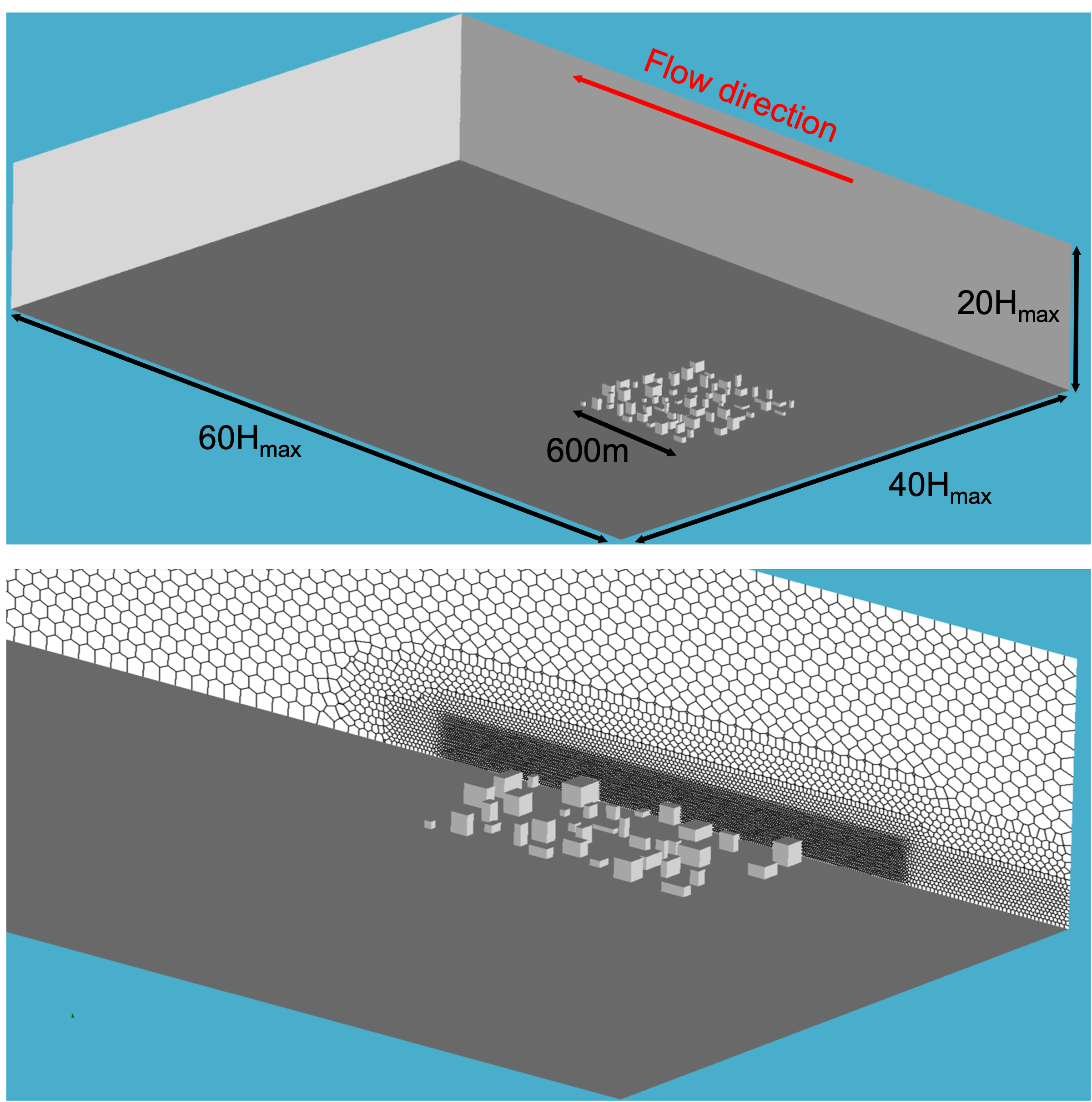}
    \caption{Computational domain (top) and zoom-in on the buildings with grid resolution (bottom).}
    \label{fig:domain_flow_pred}
    \end{figure}

Figure \ref{fig:domain_flow_pred} shows the computational domain for one of the test cases. The size of the domain is $60H_{max}$ long, $40H_{max}$ wide, and $20H_{max}$ high, where $H_{max} = 0.5$ m is the roof height of the highest building. The domain is larger than the proposed guidelines to avoid any blockage effects \citep{franke2011cost}. The computational grid is created using the CharLES mesh generator.

A grid convergence study was conducted with three different resolutions: coarse, baseline, and fine, similar to \citep{vargiemezis2024isolated}. The three meshes differ only in the grid size surrounding the buildings; the arrangement of refinement boxes and background mesh sizes remains consistent. The refinement regions surround the buildings to capture the generated turbulence and complex wake patterns that are created within the urban area. The mesh gradually coarsens as it extends away from the buildings. This process includes doubling the cell size every five layers of cells until the background cell size. Subsequently, this background cell size is maintained beyond the refinement boxes, specifically $5H_{ref}$ higher and $10H_{ref}$ downstream of the buildings. The specific details regarding background cell dimensions, minimum cell sizes, and overall number of cells are summarized in Table \ref{tab:grid_sensitivity_flow_pred}. Depending on the resolution of the grid, the time step size is adjusted to ensure a maximum Courant-Friedrichs-Lewy (CFL) number is lower than 1.0. The baseline was selected for the studies since the QoIs did not change significantly compared to the fine cases. The time step size for the baseline case is $10^{-4}$ s, corresponding to a sampling frequency of 10,000 Hz. The total duration of the simulations with the buildings is 60s. This duration corresponds to 909 $\tau_{ref}$, where $\tau_{ref}$ is the flow-through time based on the ratio of the reference building height $H_{ref} = 0.5 m$ and the speed at the same height, i.e. $W_{ref}/U_{ref} = 1/15.2 \approx $ 0.066 s. It is noted that all the QoIs were estimated after an initial burn-in period of at least 100 $\tau_{ref}$.

    \begin{table}[htpb]
    \captionof{table}{Grid resolution set-ups for the grid convergence study}
    \label{tab:grid_sensitivity_flow_pred}
    \centering
    \begin{tabular}{lccc}
    \hline
    \textbf{Cases}                      & \textbf{Coarse} & \textbf{Baseline} & \textbf{Fine} \\ \hline
    Background cell size {[}mm{]}         & 100             & 100               & 100           \\
    Smallest cell size {[}mm{]}         & 50.0             & 25.0               & 12.5           \\
    Total number of cells {[}M cells{]} & 1.2             & 2.5            & 5.1          \\ \hline
    \end{tabular}
    \end{table}

For what concerns the boundary conditions, a turbulent neutral atmospheric boundary layer is applied at the inlet with fixed roughness length $z_0 = 0.0027$,  while the two lateral boundaries are periodic, and a slip condition is applied at the top boundary. For more information regarding the inflow condition, the grid sensitivity analysis, and the LES setup, the reader is referred to \citep{vargiemezis2024isolated} since the same setup is used.

\section{Deep Neural Network method}
\subsection{Data preparation and input features}

For modeling wind around urban structures, we represent the geometry using 2D planes of the city layout. In this setup, the buildings are marked with a value of 1, while the flow field itself is marked with a value of 0. Later, the domain in the image marked with 0s will be filled with the QoIs. This method is inherently two-dimensional; however, to account for three-dimensional effects without predicting the 2D field independently, we take multiple planes at different heights. These planes are then concatenated along the channel dimension of the tensor to form a 3D representation.

\begin{figure}[htpb]
    \centering
    \includegraphics[width=\columnwidth]{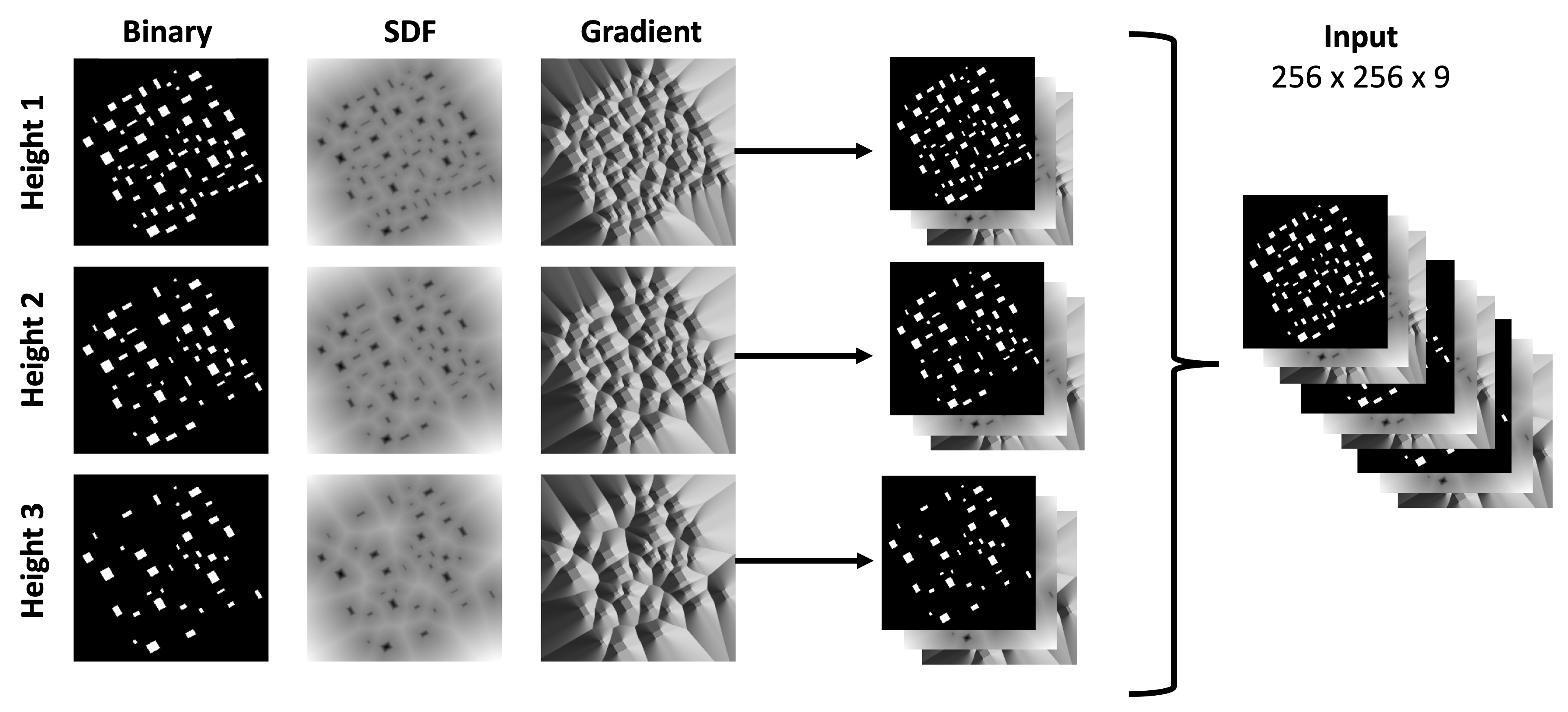}
    \caption{Geometry representation and data preparation for the U-net inputs.}
    \label{fig:data_peperation_flow_pred}
    \end{figure}

For each specified height in our model, three input images are used to include the necessary features. Figure \ref{fig:data_peperation_flow_pred} shows the formation of the input tensor using the three images for each height: 
\begin{enumerate}
    \item A binary image that separates the buildings from the flow field. In this image, buildings are represented by ones, and zeros represent the flow field.
    \item The signed distance function (SDF), which computes the Euclidean distance from the surface of the buildings. Inside the buildings, the SDF values are negative, whereas outside they are positive. This feature helps the network learn the physical boundaries and spaces of the urban area.
    \item The gradient of the SDF function, which is useful for enhancing edge detection.
\end{enumerate}

The inputs are formatted into images of size $256\times256$ pixels, with three channels per height, resulting in an input tensor of dimensions $256\times256\times9$. The output of the network predicts the two QoIs: the velocity magnitude ($U_{mag}$), and streamwise turbulence intensity ($I_u$) at three different heights. The heights are selected to be 0.1m, 0.3m, and 0.5m, but the method can be extended to more heights.

It is important to highlight that the flow direction is always from right to left along the $x$-axis, and we rotate the buildings within the domain. This is a similar approach as is usually used in CFD and wind tunnel testing, where the configuration on the turntable is typically rotated relative to the inflow direction. Additionally, keeping the flow direction fixed is beneficial for training DNNs, such as convolutional neural networks, which are not naturally invariant to rotation \citep{cohen2016group}. This consistency reduces the complexity of the learning task and can improve model generalization.

\subsection{Model architecture}
\begin{figure*}[htpb]
    \centering
    \includegraphics[width=0.74\textwidth]{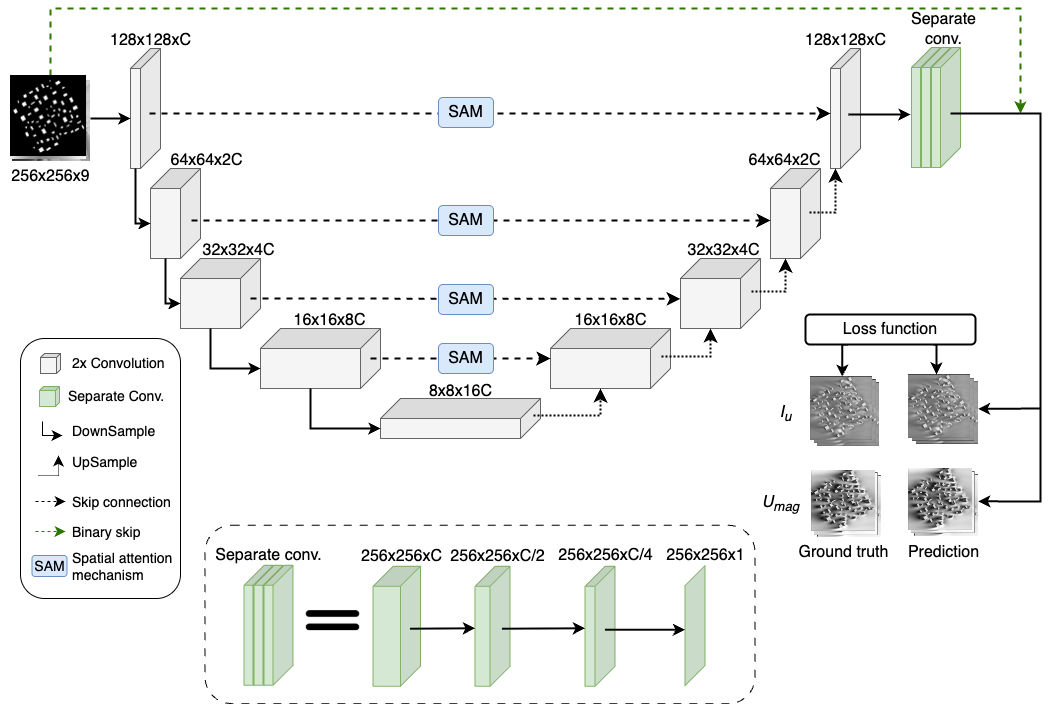}
    \caption{Model architecture for predicting velocity magnitude, $U_{mag}$, and turbulence intensity, $I_u$, at three different heights. $C$ is the number of reference channels, and it is a hyperparameter.}
    \label{fig:Unet_model_flow_pred}
    \end{figure*}

Figure \ref{fig:Unet_model_flow_pred} shows the architecture of the model. It is based on the U-net, which was originally developed for biomedical image segmentation \citep{ronneberger2015u}. The U-net has proven effective in capturing spatial features due to its encoder-decoder structure with skip connections \citep{zhai2018autoencoder}, and has shown success in various engineering applications beyond its original purpose. One notable application is flow prediction, which involves predicting the flow patterns around objects such as vehicles and airfoils \citep{thuerey2020deep, chen2022flowdnn}. The U-net capacity to capture and retain spatial information makes it particularly suited for this task.

In our U-net model, the encoder compresses the input into a lower-dimensional latent space, and the decoder reconstructs the output from this latent space. In addition, it uses skip connections between the corresponding layers in the encoder and decoder parts. These connections allow features from earlier layers to be reused during reconstruction, preserving spatial information. In particular, the encoder part consists of four convolutional blocks, each with two convolution layers and kernel size $k$, ReLU activation, and batch normalization, followed by max-pooling with a stride of 2 for downsampling. The decoder part mirrors this structure but replaces max-pooling with nearest-neighbor interpolation for upsampling, which is followed by a convolution layer.

To enhance the feature transfer across skip connections, a Spatial Attention Module (SAM) is used \citep{qin2019thundernet}. The attention mechanism ensures that critical spatial information is retained and enhanced before concatenating it with the decoder upsample. The SAM takes the form:

\begin{equation}
\mathcal{F}^{SAM} = \sigma (\text{Conv}(\mathbb{F})) \otimes \mathbb{F},
\end{equation}
where $\mathbb{F}$ is the input feature map, $\sigma$ is the sigmoid activation, and $\otimes$ is element-wise multiplication.

After decoding, the output tensor is split into multiple convolutional blocks (shown in green box), one per QoI. Each block includes a sequence of convolutional layers with a stride of 1, ReLU activation, and batch normalization. The spatial resolution of $256\times256$ is maintained, while the number of channels is halved every two layers, ending in a single-channel output image per QoI.

A binary skip connection, which is shown as a green dashed arrow, is added at the output stage. It combines the binary mask from the input with the final prediction, using element-wise addition. This makes the model to focus on learning the flow field rather than reconstructing the building surfaces, which are already given at the input.

\subsection{Loss function}
The loss function for the optimization includes three terms and is given in equation \eqref{eq:loss_flow_pred}. The first term computes the root-mean-square error (RMSE) between the predicted and the true QoI for every height/channel $H$ of the output tensor. The second term computes the RMSE of the QoI gradient magnitude based on the output image. Finally, the loss includes a regularization term that aims to reduce overfitting using $L_2$ norm on the U-net weights. The hyperparameters $\lambda_1$ and $\lambda_2$ are obtained through hyperparameter tuning. The purpose of $\lambda_1$ is to find the correct weight balance between the image loss and the gradient magnitude loss. The loss function takes the form:

\begin{align}
\mathcal{L} =\ & (1 - \lambda_1) \sum_{i=1}^{H=3} \left[ R(U_{mag}) + R(I_u) \right] \nonumber \\
& +\ \lambda_1 \sum_{i=1}^{H=3} \left[ R\left( \left\| \nabla U_{mag} \right\| \right) + R\left( \left\| \nabla I_u \right\| \right) \right] \nonumber \\
& +\ \lambda_2 \left\| W \right\|_2^2
\label{eq:loss_flow_pred}
\end{align}
where $H$ is the channel of the output tensor and corresponds to the height where the QoI was estimated. Since the QoIs are predicted at three heights, we have three channels per QoI. In addition, $R$ is the RMSE between the true and the prediction from the U-net:
    \begin{equation}
    R(x) = \sqrt{\frac{1}{n} \sum_{i=1}^n (x_i - \hat{x}_i)^2}
    \end{equation}

For what concerns the image-based gradient magnitude, it is computed using the Sobel operator, which performs a 2D spatial gradient on an image and enhances regions that correspond to edges \citep{jin2009edge}. That is, building edges, or where the gradient of the QoI is large. In practice, the 2D spatial gradient with the Sobel operator is done using a $3\times3$ convolution kernel to an image $\textbf{A}$. The gradients in the horizontal $x$ and vertical $y$ directions of the image are computed using the kernels:

\begin{align}
G_x &= 
\begin{bmatrix}
-1 & 0 & 1 \\
-2 & 0 & 2 \\
-1 & 0 & 1
\end{bmatrix}
\star \mathbf{A}, \nonumber \\
G_y &=
\begin{bmatrix}
-1 & -2 & -1 \\
0 & 0 & 0 \\
1 & 2 & 1
\end{bmatrix}
\star \mathbf{A}.
\label{eq:sobel_filters}
\end{align}
where $\star$ is the  2D signal processing convolution operation. Finally, the gradient magnitude is computed with:
\begin{equation}
\left \| \nabla \right \| = \sqrt{G_x^2 + G_y^2}.
\end{equation}

\subsection{Hyperparameter tuning}
The hyperparameters were optimized through a stochastic grid search \citep{bergstra2012random}. Detailed information on the hyperparameters, including their tuning ranges and optimal values, can be found in Table \ref{tab:hyperparameters_flow_pred}. The model was trained for 500 epochs using minibatch gradient descent with the Adam optimizer \citep{kingma2014adam}.

The dataset includes 252 cases, which were divided into training, development (dev), and test sets in an 80:10:10 ratio. This partitioning resulted in 202 cases for training, 25 for dev, and 25 for testing. Subsequently, we applied data augmentation to each set separately by flipping each image vertically along the $y$-axis, which corresponds to the spanwise direction of the mean wind speed. This data augmentation technique is appropriate for our approach, as we consider wind directions ranging from $0^{\circ}-90^{\circ}$, with the flow consistently aligned horizontally (from right to left), along the $x$-axis. After the data augmentation, the train:dev:test ratio is the same, and the total number of cases are 404:50:50.

The training was done using one RTX3090 GPU, and the hyperparameter tuning takes $\sim 72$ hours. The test set was used for the assessment and the accuracy of the model in predicting the QoIs. The evaluation of the model, along with the results are presented in the next chapter.

\begin{table}[htbp]
\centering
\caption{Hyperparameter Tuning}
\small
\begin{tabular}{@{}lcc@{}}
\toprule
\textbf{Hyperparameter} & \textbf{Tuning Range} & \textbf{Optimum} \\
\midrule
Channels, $C$ & \{4, 8, 16, 32, 64, 128\} & 128 \\
Kernel size, $k$ & \{3, 5, 7\} & 5 \\
Gradient weight, $\lambda_1$ & $[10^{-7},\,10^{-1}]$ & $10^{-3}$ \\
Reg. weight, $\lambda_2$ & $[10^{-7},\,10^{-1}]$ & $7{\times}10^{-5}$ \\
Batch size & \{4, 8, 16, 32\} & 16 \\
Learning rate, $L_r$ & $[10^{-5},\,10^{-2}]$ & $5{\times}10^{-4}$ \\
\bottomrule
\end{tabular}
\label{tab:hyperparameters_flow_pred}
\end{table}

\section{Results and discussion}
In this chapter, we present the results of the U-net predictions for the two QoIs: 1) mean velocity magnitude and 2) streamwise turbulence intensity. For reference, a single evaluation using the U-net model takes time in the order of $\mathcal{O}1$ seconds on one GPU, compared to $\mathcal{O}10$ hours for an LES evaluation on 32 CPUs. 

The U-net model predictions are compared against the ground truth provided by the LES dataset. The evaluation includes a test set of 50 cases, though the focus is on two representative cases to illustrate the overall performance. These cases include different city layouts, one at $0^{\circ}$ and the other at $45^{\circ}$. For both representative wind directions, we begin by presenting contour plots of the QoIs at three different heights. Following this, we provide line plots of the same QoIs at various vertical lines across the domain. For a more quantitative comparison, we include histograms of the prediction errors. Finally, we assess the overall performance of the model using three different metrics applied to the entire test set: hit rate, root-mean-square error, and mean relative error.

\subsection{Overview of evaluation metrics}
To assess the accuracy of the U-net model, we consider four primary metrics: the signed absolute error, the hit rate, the normalized root-mean-square error (NRMSE), and the mean relative error (MRE).

\noindent\textbf{Signed Absolute Error}. For histogram analysis, we define the signed absolute error $\Delta_i$ as:
\begin{equation}
    \Delta_i = \frac{|Q_i^T - Q_i^M|}{\text{sign}(|Q_i^T| - |Q_i^M|)},
\end{equation}
where $Q_i^T$ and $Q_i^M$ are the true and predicted values of the quantity of interest at pixel $i$.

\noindent\textbf{Hit Rate}. This metric represents the percentage of pixels for which the U-net prediction falls within a predefined error threshold relative to the LES ground truth. Specifically, for velocity magnitude $U_{mag}$, the threshold is set at $\pm 15\%$ of the inlet velocity at 10 m height, while for turbulence intensity $I_u$, it is $\pm 15\%$ of the inlet turbulence intensity at the same reference height.

\noindent\textbf{Normalized Root-Mean-Square Error (NRMSE)}. The NRMSE quantifies the average magnitude of the prediction errors, normalized by the range of ground truth values:
\begin{equation}
    NRMSE = \sqrt{ \frac{1}{N} \cdot \frac{\sum_{i=1}^N (Q_i^T - Q_i^M)^2}{|Q_{max}^T - Q_{min}^T|} },
\end{equation}

\noindent\textbf{Mean Relative Error (MRE)}. The MRE measures the average relative difference between predictions and ground truth, expressed as a percentage:
\begin{equation}
    MRE = \frac{1}{N} \sum_{i=1}^{N} \left( \frac{|Q_i^T - Q_i^M|}{|Q_i^T|} \right) \times 100\%.
\end{equation}

\subsection{Flow predictions at $0^{\circ}$ wind incidence}
\subsubsection{Contour plots of $U_{mag}$ and $I_u$}
    \begin{figure}[htbp]
        \centering
        \includegraphics[width=\linewidth]{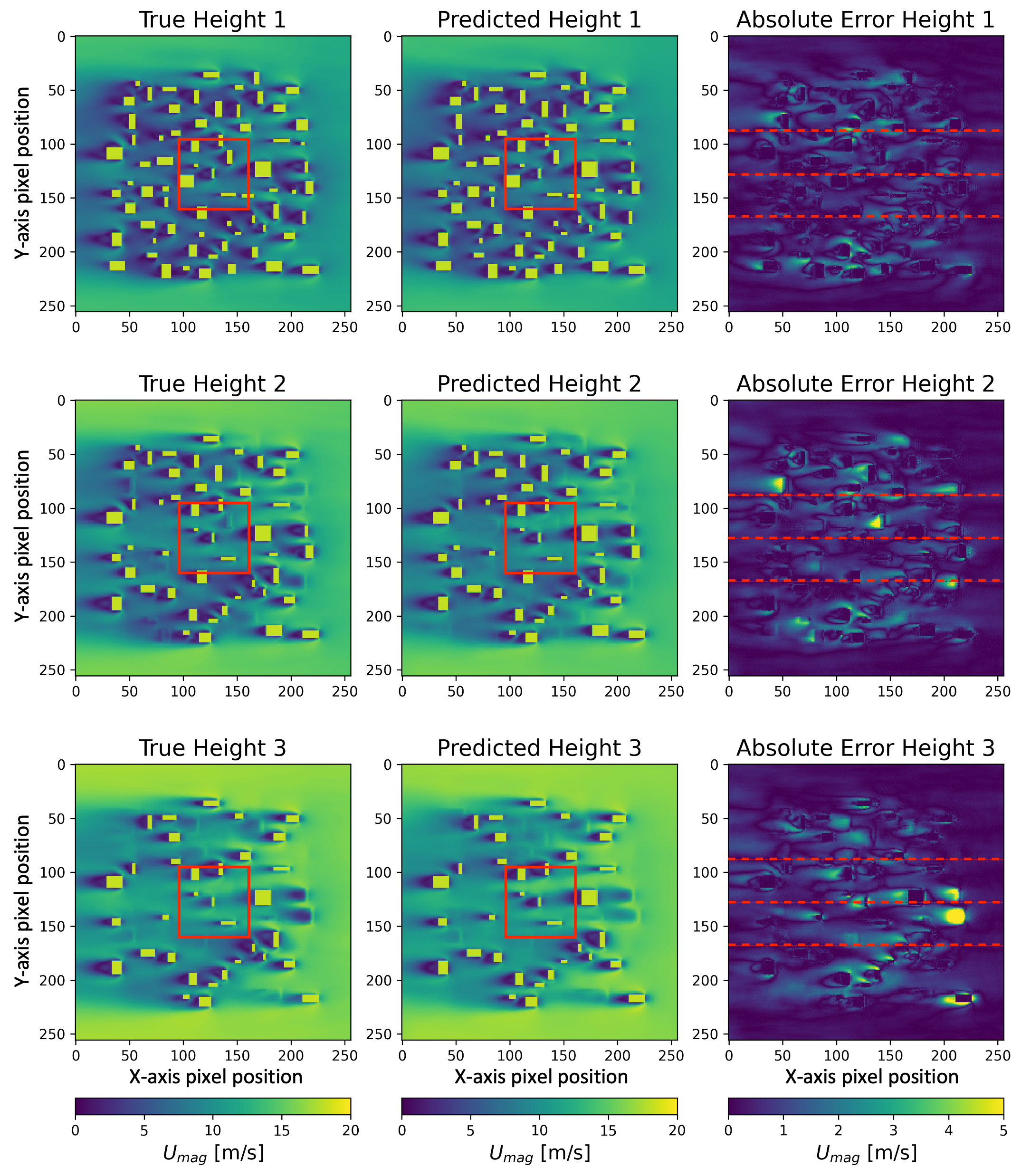}
        \caption{Velocity magnitude \(U_{mag}\) at \(0^{\circ}\) wind incidence. Left: LES ground truth, middle: U-net prediction, right: absolute error. Height 1, 2, and 3 correspond to 0.1m, 0.3m, and 0.5m.}
        \label{fig:contour_mag2_0deg_flow_pred}
    \end{figure}
    
    \begin{figure}[htbp]
        \centering
        \includegraphics[width=\linewidth]{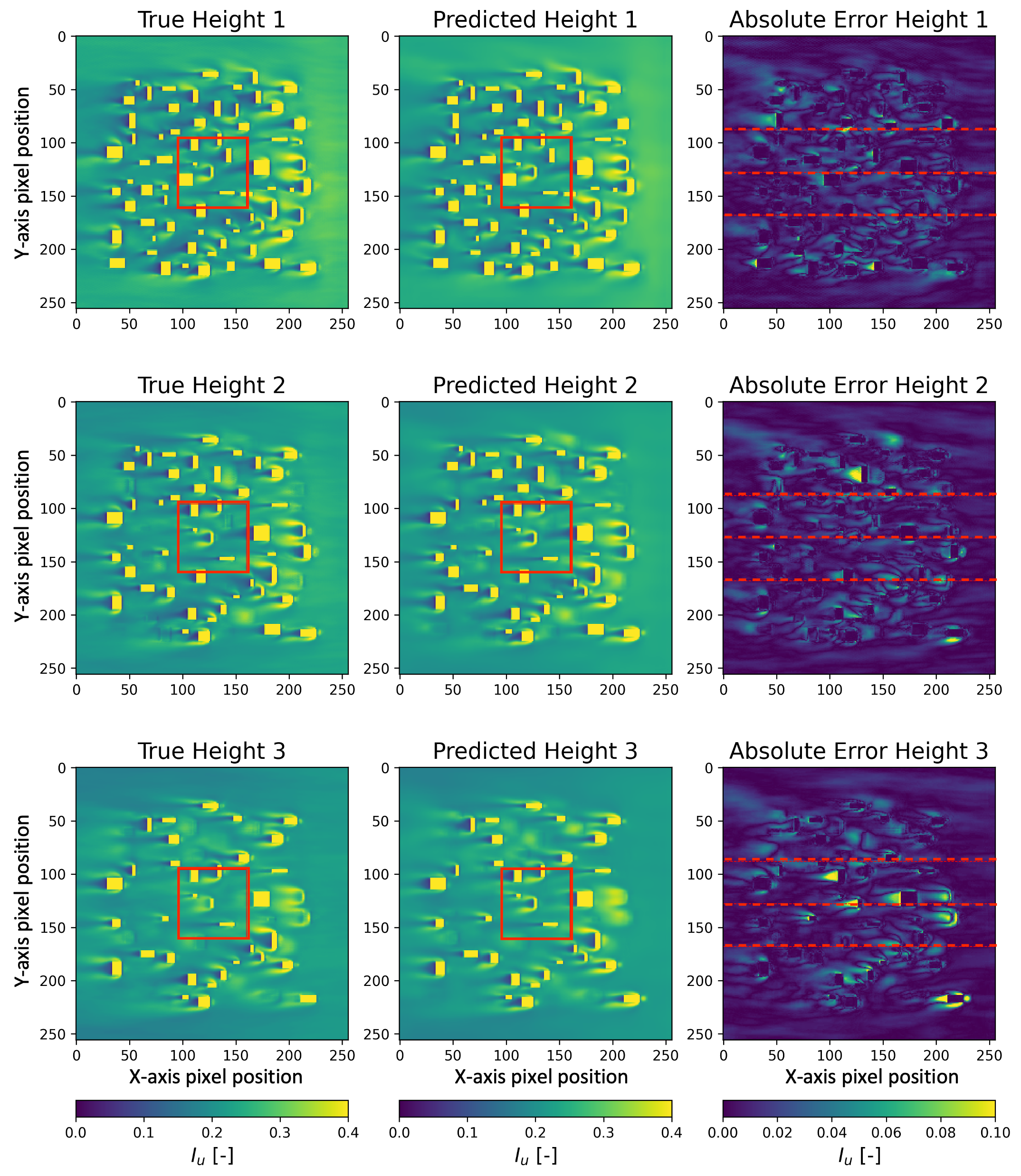}
        \caption{Turbulence intensity \(I_u\) at \(0^{\circ}\) wind incidence. Left: LES ground truth, middle: U-net prediction, right: absolute error. Height 1, 2, and 3 correspond to 0.1m, 0.3m, and 0.5m.}
        \label{fig:contour_Iu2_0deg_flow_pred}
    \end{figure}

    Figure \ref{fig:contour_mag2_0deg_flow_pred} shows the contour plot comparison of the velocity magnitude, while Figure \ref{fig:contour_Iu2_0deg_flow_pred} shows the comparison of the turbulence intensity. The red dashed lines correspond to the locations where the line plots are plotted in the next section, and the red box corresponds to the region of interest (RoI). Often, instead of focusing on the whole urban area, we want to study the flow field in a RoI, around a target building at the center of the city. This approach is usually followed in wind engineering applications. A comparison of the accuracy in predicting the RoI compared to the full flow field prediction is given in the next sections.
    
    In the velocity predictions presented in Figure~\ref{fig:contour_mag2_0deg_flow_pred}, it is observed that the U-net model is able to predict accurately the flow field within the urban canopy across all considered heights. Notably, the U-net has the ability to infer three-dimensional flow effects, such as wake formation and flow separation induced by variations in building height, despite being provided with only two-dimensional representations of the urban geometry. This behavior is particularly visible at Height 3, where the influence of lower buildings on the flow field is captured. However, this behavior is limited, which can be seen from the largest discrepancies in velocity magnitude that are also observed at this height, with absolute errors reaching up to 5 m/s. Including additional vertical information, such as more height slices of the urban canopy, may help mitigate these limitations and improve the ability to capture three-dimensional flow structures more accurately. 

    The predictions of turbulence intensity in Figure \ref{fig:contour_Iu2_0deg_flow_pred} show a similar behavior with a high level of agreement, and some locations reach an absolute error of up to 0.1. These discrepancies are mostly outside of the RoI. It is also evident that the U-net can predict the three-dimensional flow patterns at higher heights, with some visible discrepancies. These discrepancies can also be mitigated by taking into account more heights of the urban canopy.
    
\subsubsection{Line plots of $U_{mag}$ and $I_u$}
\begin{figure*}[htbp]
    \centering
    \begin{subfigure}[t]{0.8\textwidth}
        \includegraphics[width=\textwidth]{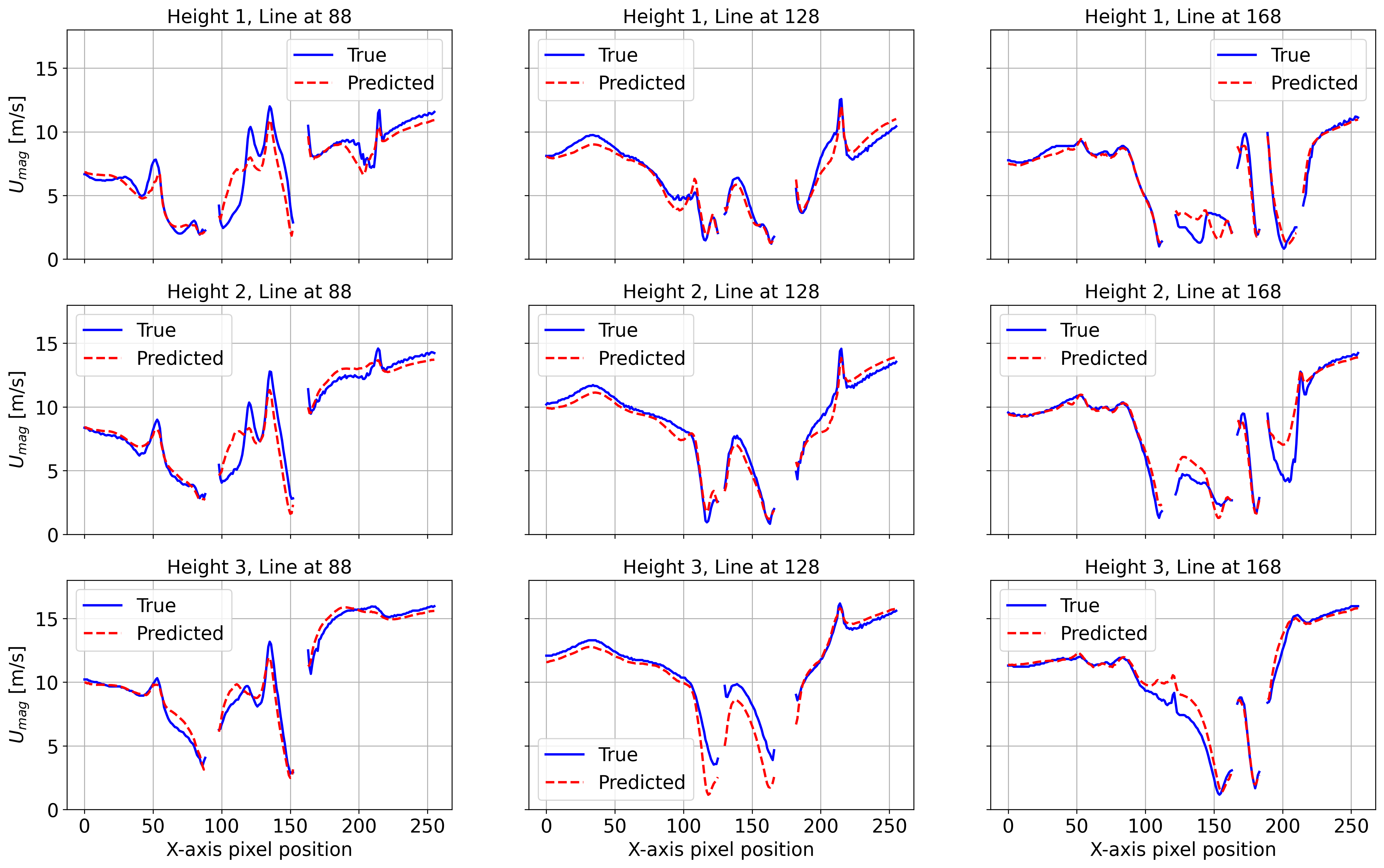}
        \caption{$U_{mag}$ comparison}
        \label{fig:lineplots_magu_deg0_flow_pred}
    \end{subfigure}
    
    \vspace{1em}  
    
    \begin{subfigure}[t]{0.8\textwidth}
        \includegraphics[width=\textwidth]{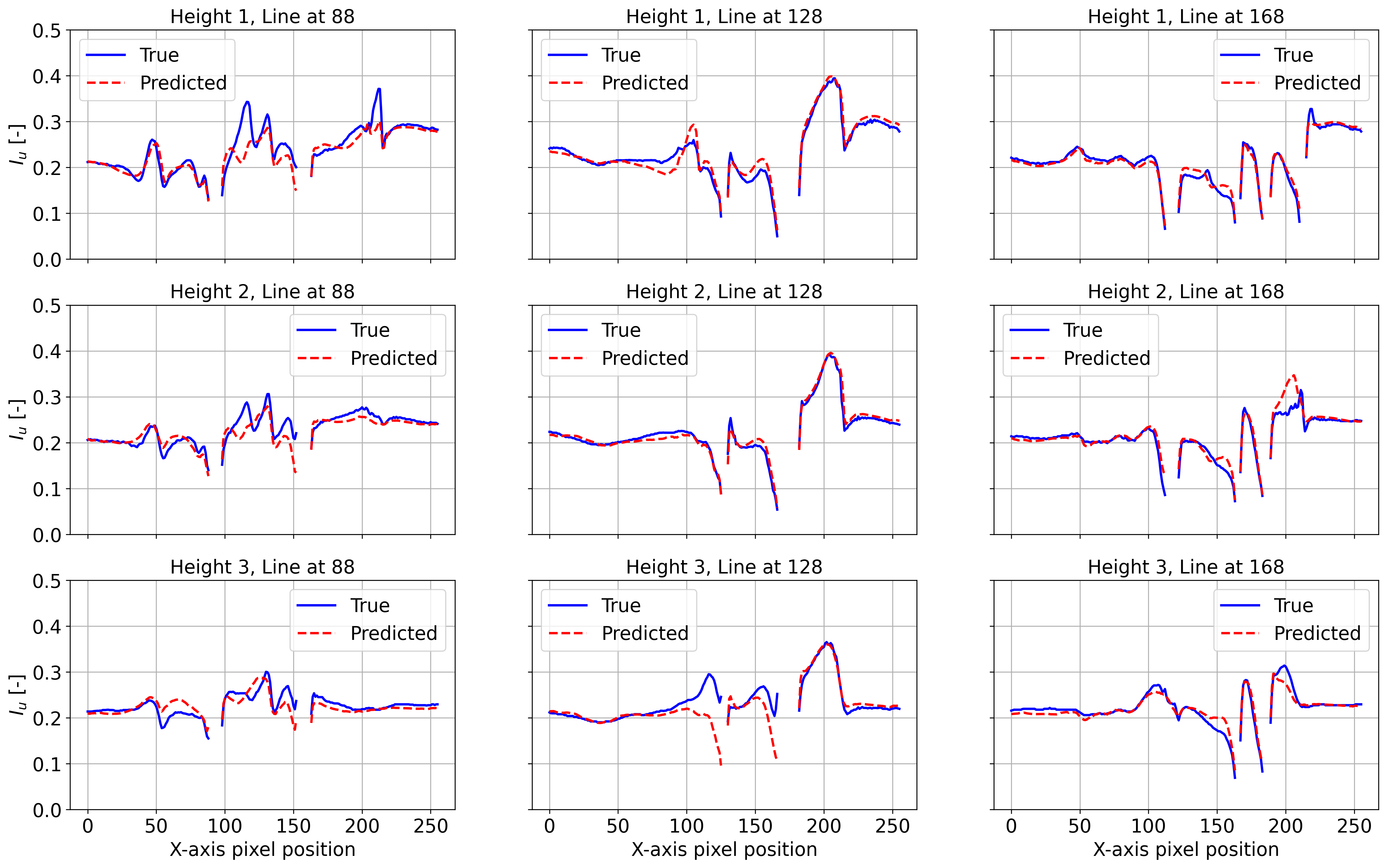}
        \caption{$I_{u}$ comparison}
        \label{fig:lineplots_Iu_deg0_flow_pred}
    \end{subfigure}
    \caption{Line plot comparisons of $U_{mag}$ and $I_u$ at $0^{\circ}$ wind incidence.}
    \label{fig:lineplots_deg0_flow_pred}
\end{figure*}

    For a more quantitative comparison, Figures \ref{fig:lineplots_magu_deg0_flow_pred} and \ref{fig:lineplots_Iu_deg0_flow_pred} show the profiles of the velocity magnitude and turbulence intensity predictions at three lines that span the domain in the streamwise $x$-direction. The $x$‑axis is spanned in the opposite direction of the flow, where pixel 0 lies downstream and pixel 250 upstream. The U-net predictions are shown with dashed red lines, while the LES ground truth is shown with solid blue lines.
    
    Overall, the \(U_{mag}\) and \(I_u\) profiles show excellent agreement at all heights, with only minor discrepancies in the wake regions of buildings. For instance, in Figure \ref{fig:lineplots_magu_deg0_flow_pred}, at Height 3 and along the centerline at \(y = 128\), the U-net underpredicts the velocity between $x = 100$ and 175 by up to 1–2 m/s, which is roughly 10–15\% of the local wind speed, but correctly captures the downstream recovery trend. Upstream, at $x > 200$ and in the free‑stream region at $x < 50$, errors drop to under 5\% of the mean flow, showing that the network has learned the bulk profiles of the flow field.

    Regarding \(I_u\) in Figure \ref{fig:lineplots_Iu_deg0_flow_pred}, similar patterns are observed with excellent overall agreement and localized discrepancies in wake regions. The U-net reproduces the \(I_u\) peaks downstream of buildings with good accuracy, correctly identifying the locations of turbulence production in shear layers and building wakes, though it slightly underpredicts peak values in some complex wake regions. Some discrepancies are observed at the same location as in the \(U_{mag}\); at Height 3 of $y = 128$, and at $x$-axis pixel values between 100 and 175. The U-net underpredicts \(I_u\) by approximately 0.1 in these regions, representing roughly 20\% of the local turbulence intensity values. This location corresponds to a building located at $(x,y) = (128, 128)$ in the contour plot of Figure \ref{fig:contour_Iu2_0deg_flow_pred}, where the U-net model does not accurately capture the full complexity of the wake region.
    
    The height-dependent performance shows the cause of these discrepancies and provides insight into the limitations of this model. At Heights 1 and 2, where the flow is more constrained by the urban canopy and exhibits 2D characteristics, the U-net predictions show better agreement with the LES ground truth. However, at Height 3, where three-dimensional effects become more important due to varying building heights and increased vertical flow characteristics, the discrepancies become more obvious for both $U_{mag}$ and $I_u$ predictions. This reduced performance at higher heights shows the inherent challenge of inferring 3D flow structures from 2D geometric representations, particularly in regions where vertical flow components play significant roles in determining the local flow field. 
   
\subsubsection{Absolute error histogram comparison}
\begin{figure*}[htpb]
    \centering
    \begin{subfigure}[t]{0.82\textwidth}
        \includegraphics[width=\textwidth]{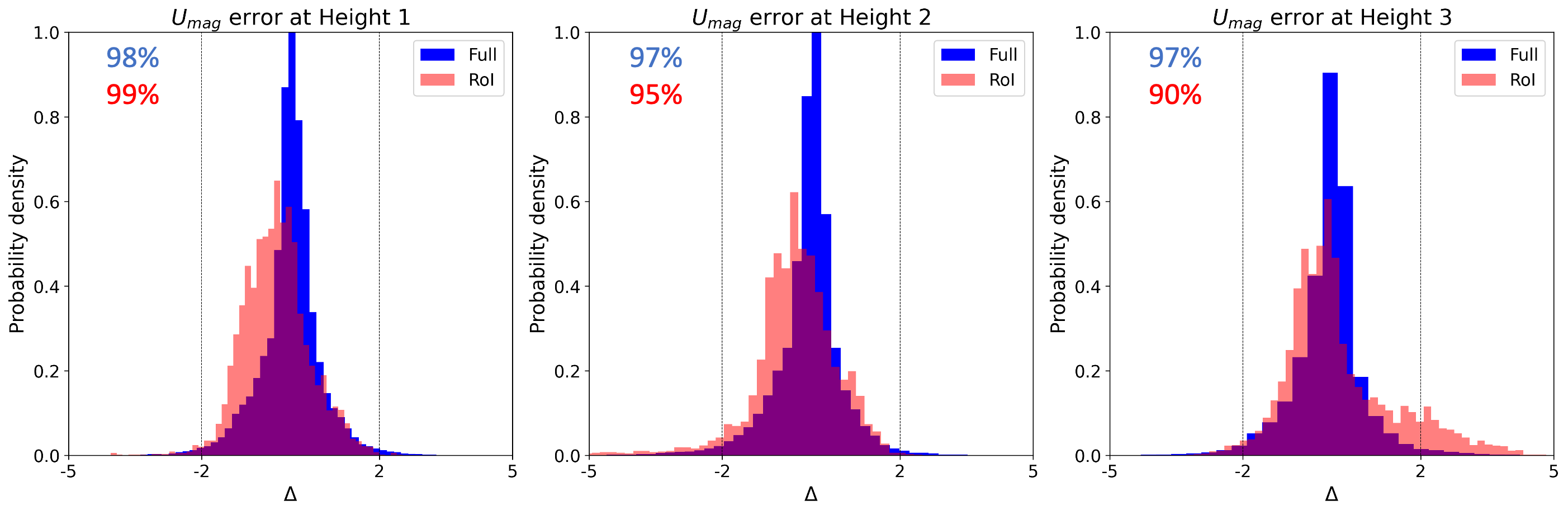}
        \caption{$U_{mag}$ comparison}
        \label{fig:histogram_magu_deg0_flow_pred}
    \end{subfigure}
    
    \vspace{1em}
    
    \begin{subfigure}[t]{0.82\textwidth}
        \includegraphics[width=\textwidth]{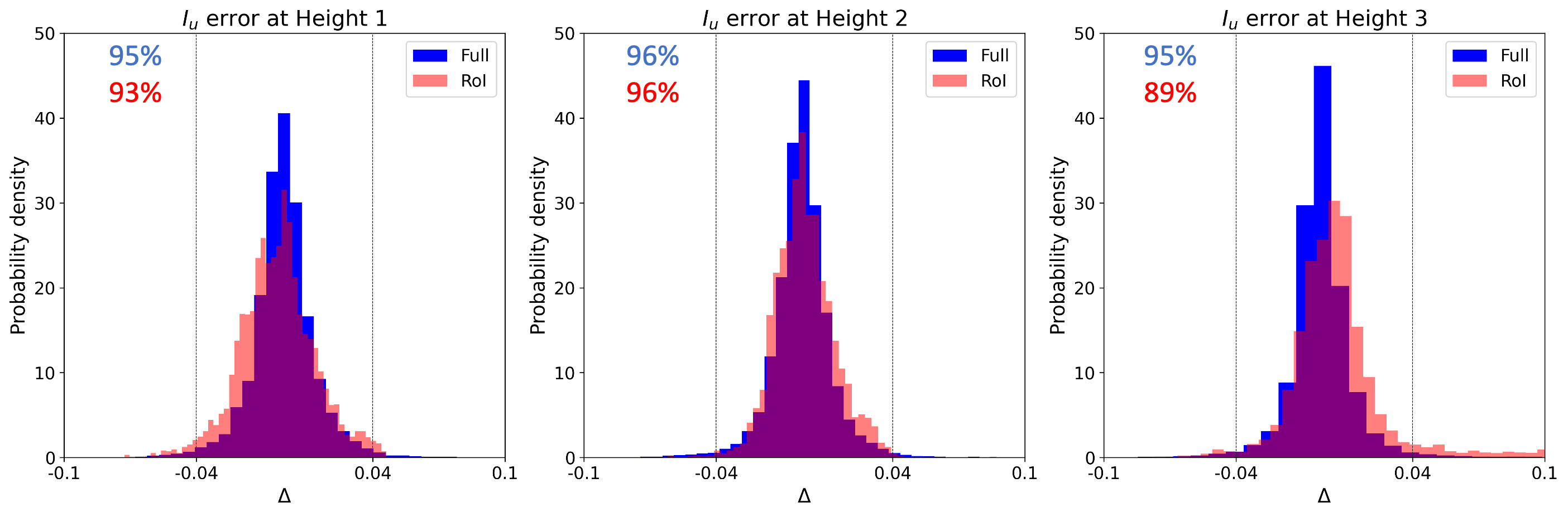}
        \caption{$I_{u}$ comparison}
        \label{fig:histogram_Iu_deg0_flow_pred}
    \end{subfigure}
    \caption{Error histogram at $0^{\circ}$ wind incidence.}
    \label{fig:histogram_deg0_flow_pred}
\end{figure*}

    The histograms provided in Figures \ref{fig:histogram_magu_deg0_flow_pred} and \ref{fig:histogram_Iu_deg0_flow_pred} show the signed absolute error, $\Delta_i$, distributions of the U-net predictions against the LES ground truth for the flow field in the city models. The histograms are compared for both the full domain and the RoI at the center of the city. Hit rates are also given as percentages. The results are presented at the three different heights.

    The histograms show that the U-net prediction error for the full domain in blue and the RoI in red are centered around zero, indicitating unbiased predictions overall. However, the distributions show slight asymmetries, specifically at Height 3, where there is a small shift toward positive values, indicating underprediction for both \(U_{mag}\) and \(I_u\) by the U-net model. This aligns with the wake region underpredictions by U-net that were observed in the line plots behind the building located at $(x,y) = (128, 128)$ in Figure \ref{fig:lineplots_magu_deg0_flow_pred}.

    The height-dependent performance can also be observed by the histograms. At Height 1, the prediction error for \(U_{mag}\) and \(I_u\) are mostly within the $\pm 15\%$ thresholds, showing high accuracy. This is confirmed by the hit rate values of 98\% (full) and 99\% (RoI) for \(U_{mag}\) and similar values for \(I_u\), with 95\% ad 93\%, respectively. At Height 2, the prediction errors remain low and mostly centered around zero, with hit rate values of 97\% (full) and 95\% (RoI) for \(U_{mag}\), and 96\% for \(I_u\) in both full domain and RoI. At Height 3, while prediction error remain low, especially for \(U_{mag}\), the hit rates slightly decrease to 97\% for the full domain and 90\% for RoI. Similarlry, the hit rates for \(I_u\) are 95\% for the full domain and 89\% for the RoI. These values confirm the slight decrease in predictive accuracy at higher heights.

    The comparison between the full domain and the RoI is also important to understand the predictive capabilities of the model inside the urban area. The higher hit rate observed for the full flow field prediction can be attributed to the ability of the U-net to accurately infer the uniform flow outside the urban canopy, where building wakes are absent. The inclusion of the uniform outer flow contributes a large number of correctly predicted points, thereby increasing the overall hit rate. The lower hit rate of RoI, especially at higher heights, show the inherent challenge of inferring 3D flow structures from 2D geometric representations.

\subsection{Flow predictions at $45^{\circ}$ wind incidence}
\subsubsection{Contour plots of $U_{mag}$ and $I_u$}
    \begin{figure}[htbp]
        \centering
        \includegraphics[width=\linewidth]{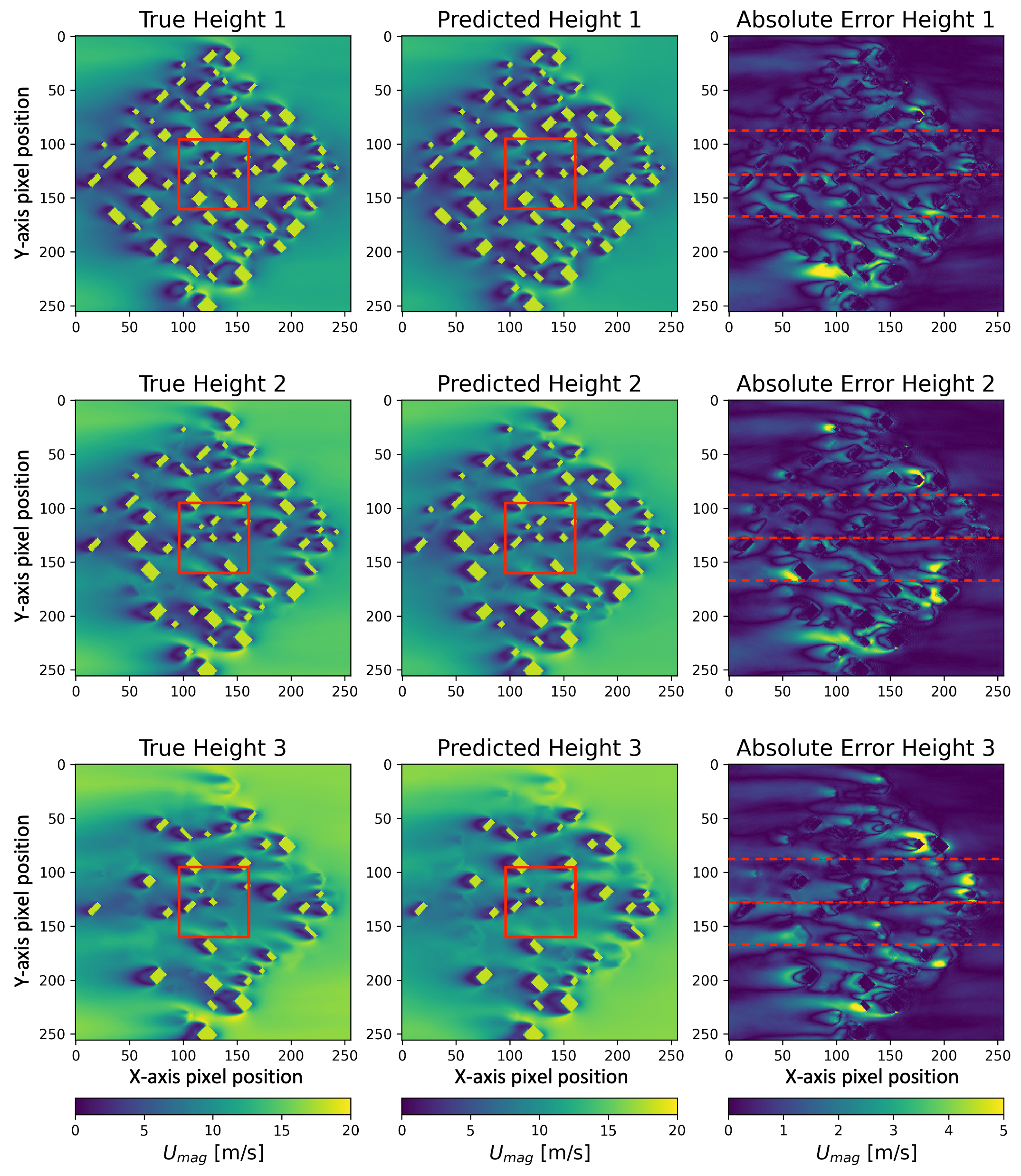}
        \caption{Velocity magnitude \(U_{mag}\) at \(45^{\circ}\) wind incidence. Left: LES ground truth, middle: U-net prediction, right: absolute error. Height 1, 2, and 3 correspond to 0.1m, 0.3m, and 0.5m.}
        \label{fig:contour_mag2_45deg_flow_pred}
    \end{figure}
    
    \begin{figure}[htbp]
        \centering
        \includegraphics[width=\linewidth]{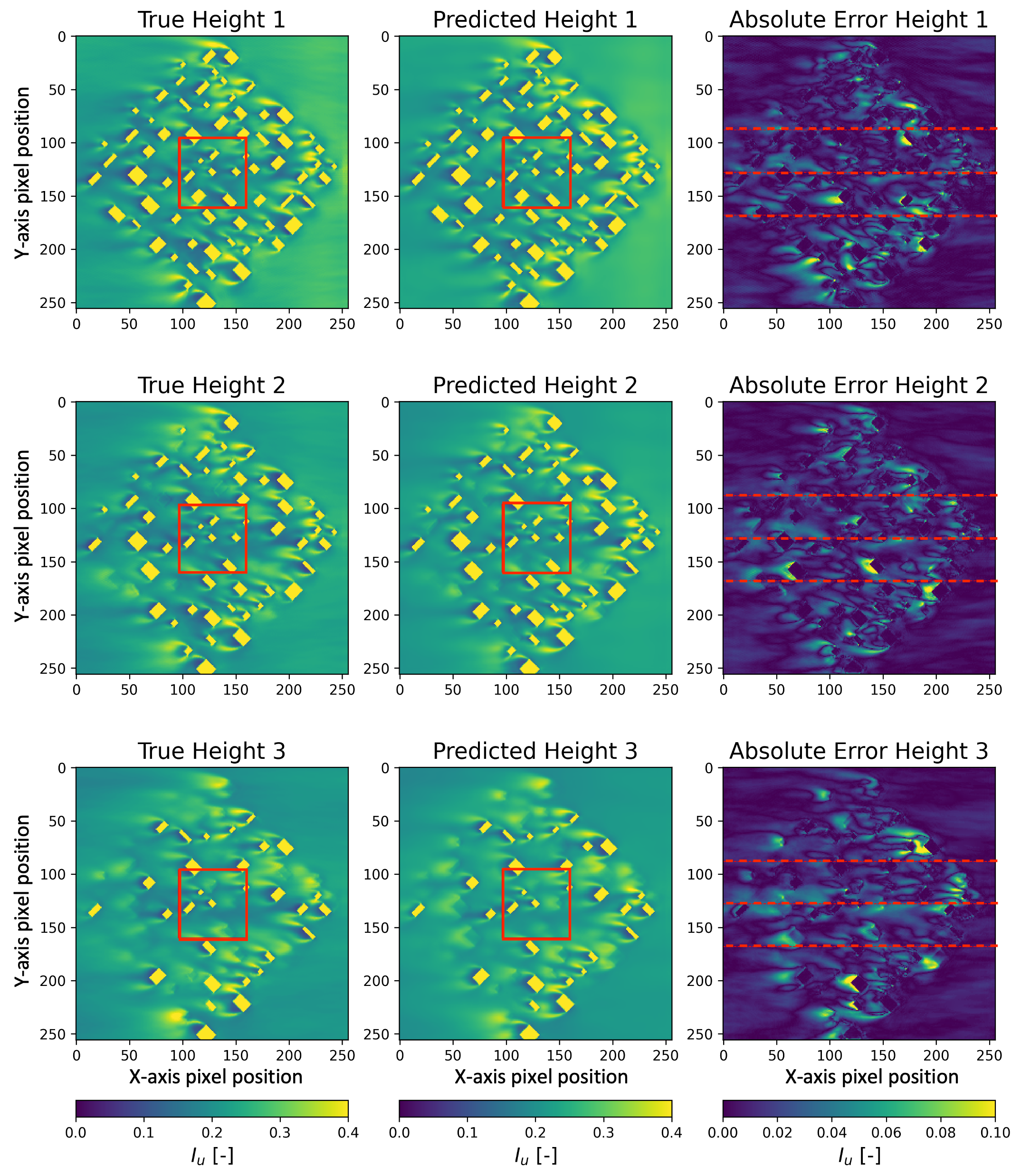}
        \caption{Turbulence intensity \(I_u\) at \(45^{\circ}\) wind incidence. Left: LES ground truth, middle: U-net prediction, right: absolute error. Height 1, 2, and 3 correspond to 0.1m, 0.3m, and 0.5m.}
        \label{fig:contour_Iu2_45deg_flow_pred}
    \end{figure}

    In Figure \ref{fig:contour_mag2_45deg_flow_pred}, the velocity magnitude prediction at $45^{\circ}$ shows good agreement at all three heights, with only some local regions with relatively large errors. Notably, at all heights, there is a region at the pixel locations $(x,y) = (120, 225)$ where the building wake is not captured correctly, and the predictions reach an absolute error of 5 m/s locally. A similar conclusion can be drawn at the location $(x,y) = (180, 60)$. Both locations correspond to the upstream edges of the urban canopy, while at the RoI, the U-net model shows a better qualitative agreement with the ground truth. In addition, similar to the $0^\circ$ case, the errors are larger at higher heights.
    
    The predictions of turbulence intensity at $45^{\circ}$, shown in Figure \ref{fig:contour_Iu2_45deg_flow_pred}, show good qualitative agreement with the ground truth. The locations of high absolute error match the locations observed in the velocity predictions, and in the RoI, the predictions are more accurate.  

    It is important to note that, in most cases, the flow patterns, such as wake formation and flow separation, are accurately captured for both QoIs. The primary discrepancies come from differences in the magnitude of each QoI, with some locations experiencing underprediction or overprediction. These discrepancies will become more apparent in the line plots presented in the next section.

\subsubsection{Line plots of $U_{mag}$ and $I_u$}
\begin{figure*}[htpb]
    \centering
    \begin{subfigure}[t]{0.8\textwidth}
        \includegraphics[width=\textwidth]{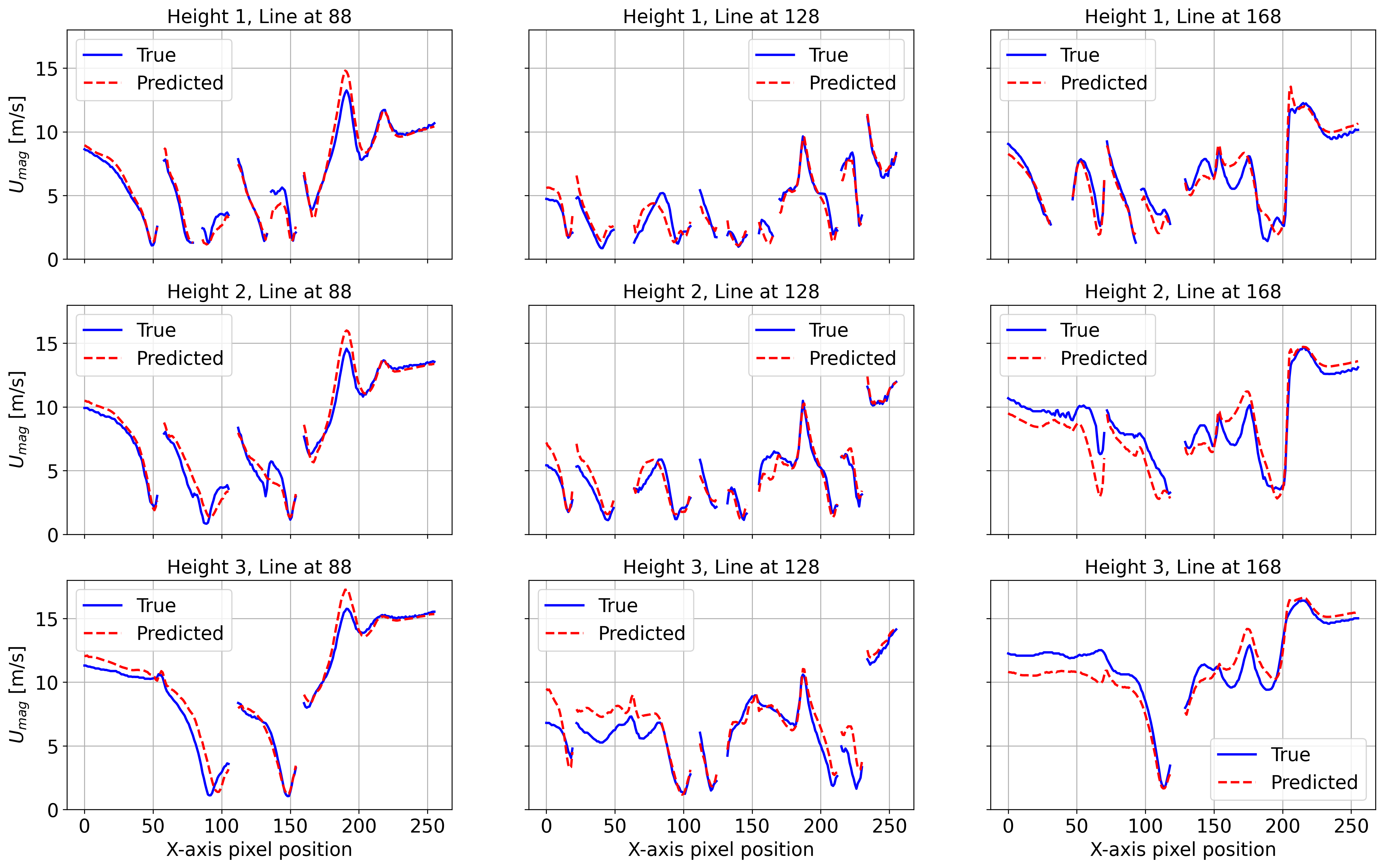}
        \caption{$U_{mag}$ comparison}
        \label{fig:lineplots_magu_deg45_flow_pred}
    \end{subfigure}
    
    \vspace{1em}
    
    \begin{subfigure}[t]{0.8\textwidth}
        \includegraphics[width=\textwidth]{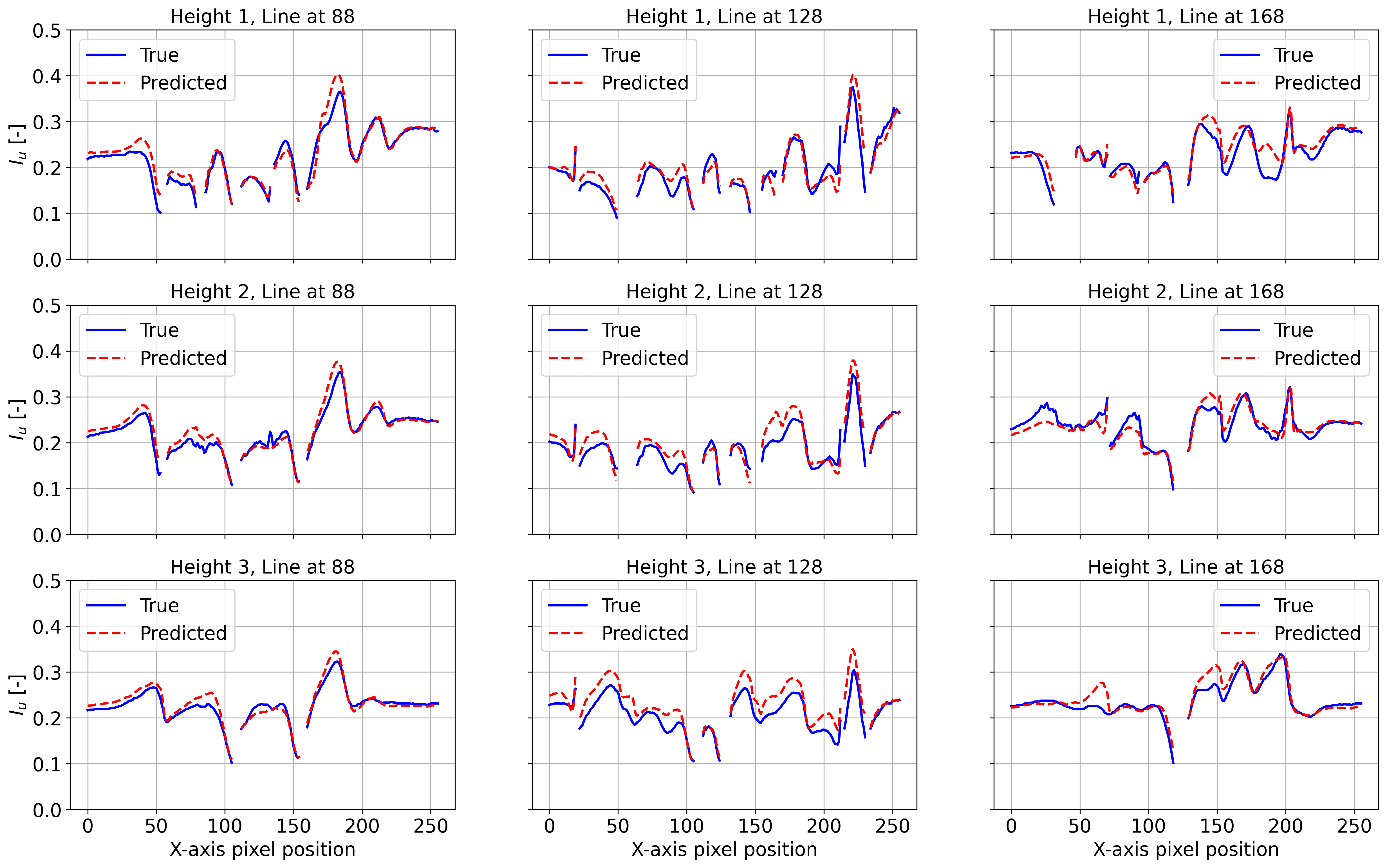}
        \caption{$I_{u}$ comparison}
        \label{fig:lineplots_Iu_deg45_flow_pred}
    \end{subfigure}
    \caption{Line plot comparisons of $U_{mag}$ and $I_u$ at $45^{\circ}$ wind incidence.}
    \label{fig:lineplots_deg45_flow_pred}
\end{figure*}

    Figures \ref{fig:lineplots_magu_deg45_flow_pred} and \ref{fig:lineplots_Iu_deg45_flow_pred} show the profiles of the velocity magnitude and turbulence intensity at three lines that span the domain and at the three heights. The U-net predictions is shown in dashed red lines, while the LES is shown with solid blue lines.

    For the \(U_{mag}\) predictions in Figure~\ref{fig:lineplots_magu_deg45_flow_pred}, the agreement is excellent at Heights 1 and 2. However, consistent with the $0^{\circ}$ case, the largest discrepancies are observed at Height 3, where three-dimensional flow effects become more dominant. For instance, along the line at \(y=128\), the U-net overpredicts the velocity in the wake region between $x$-pixel values 0 and 75 by up to 3~m/s, a relative error of approximately 25\% of the local velocity. A similar, though less pronounced, underprediction of approximately 2~m/s, a relative error of approximately 15\% of the local velocity, is visible along the line at \(y=168\) in the wake at $x$-pixel values between 0 and 75. This further highlights the limitations of fully inferring complex 3D flow structures from the provided 2D inputs.

    Regarding the \(I_u\) predictions in Figure~\ref{fig:lineplots_Iu_deg45_flow_pred}, the U-net again predicts accurately the flow field, inferring the locations of turbulence production, which are identified by the sharp peaks in the profiles. While the location of these peaks is well-predicted, the model tends to slightly overpredict their magnitude, particularly at higher heights. This is observable at Height 3 along the line \(y=128\), where the U-net overpredicts the peak turbulence intensity by approximately 0.05, a relative error of nearly 15\%. Despite these localized discrepancies in magnitude, the predictions follow the overall trend, indicating the model has learned the regions of high turbulence downstream of the buildings.

    Overall, the profiles for both \(U_{mag}\) and \(I_u\) show high accuracy, confirming that the U-net model can effectively capture the more complex flow patterns generated by the angled buildings with respect to the wind direction. 
    
\subsubsection{Absolute error histogram comparison}
\begin{figure*}[htpb]
    \centering
    \begin{subfigure}[t]{0.82\textwidth}
        \includegraphics[width=\textwidth]{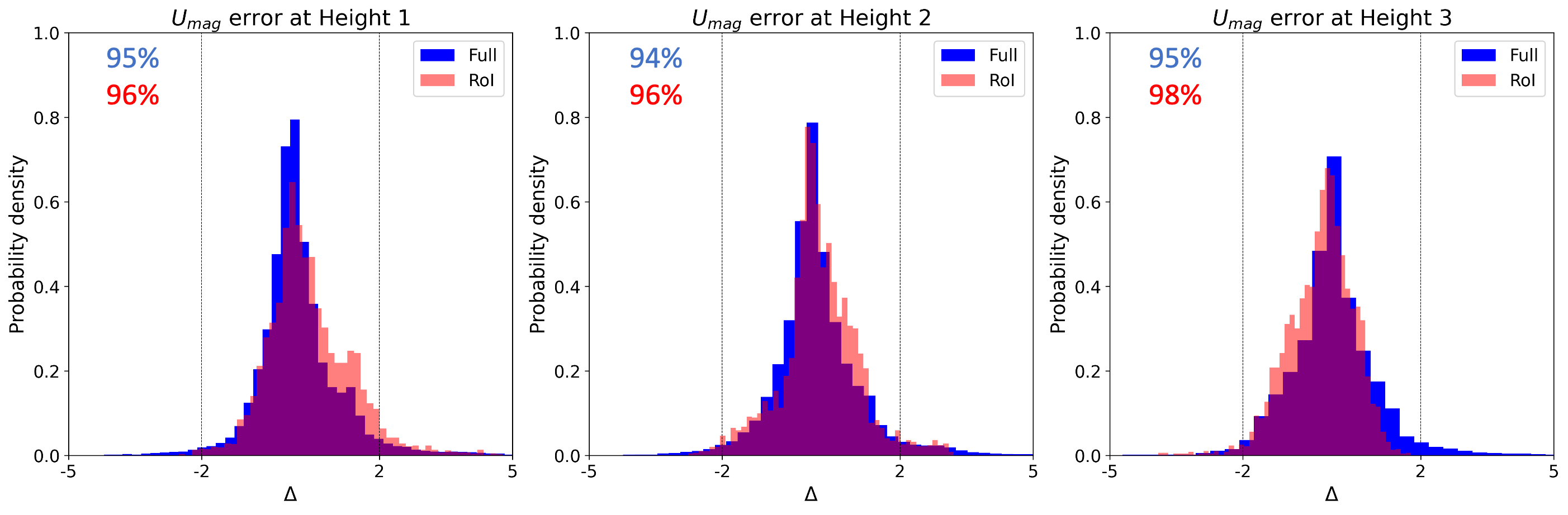}
        \caption{$U_{mag}$ comparison}
        \label{fig:histogram_magu_deg45_flow_pred}
    \end{subfigure}
    
    \vspace{1em}
    
    \begin{subfigure}[t]{0.82\textwidth}
        \includegraphics[width=\textwidth]{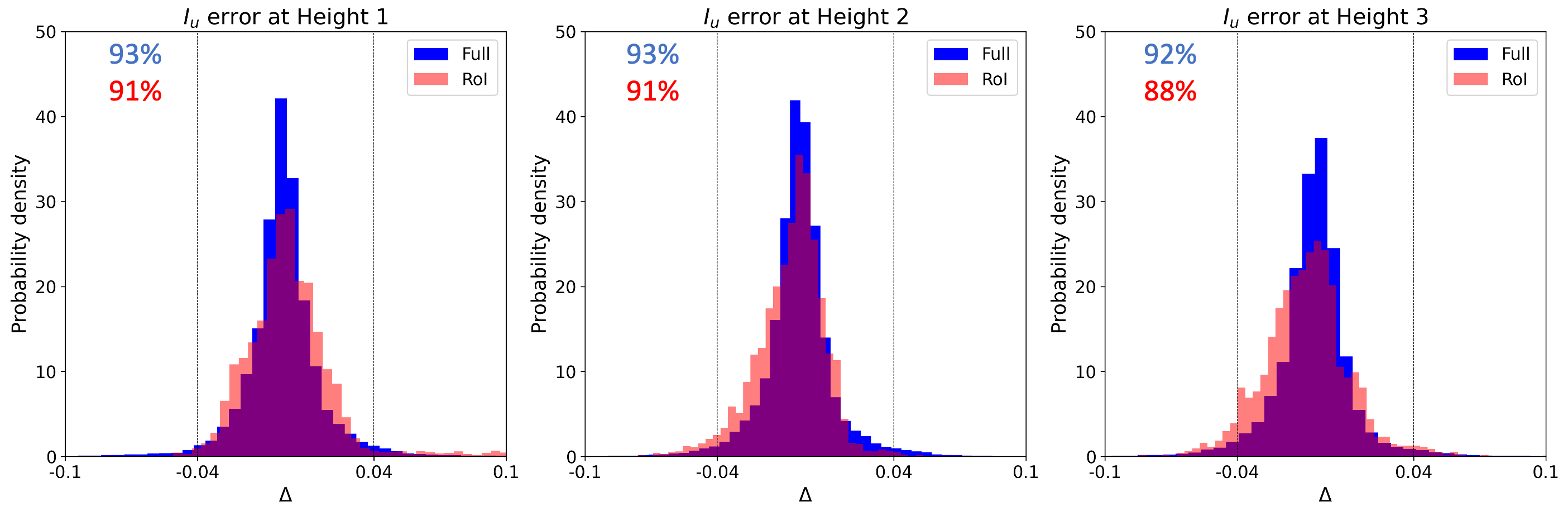}
        \caption{$I_{u}$ comparison}
        \label{fig:histogram_Iu_deg45_flow_pred}
    \end{subfigure}
    \caption{Error histogram at $45^{\circ}$ wind incidence.}
    \label{fig:histogram_deg45_flow_pred}
\end{figure*}

    Figure \ref{fig:histogram_magu_deg45_flow_pred} and Figure \ref{fig:histogram_Iu_deg45_flow_pred} show the histograms of the signed absolute error, $\Delta_i$, for the U-net predictions at the 45° wind incidence. The distributions compare the performance on the full domain (blue) against the RoI (red) at the three different heights, with the corresponding hit rates.

    Similarlry to the $0^{\circ}$ case, the U-net prediction errors are centered around zero, showing unbiased prediction overall, but the distributions appear broader. For $U_{mag}$ predictions, the hit rates for both the RoI and the full flow field are comparable, exceeding 94\% at all heights. Notably, for this wind direction, the hit rate within the RoI is consistently higher than or equal to that of the full domain, reaching 98\% at Height 3. A slight positive skew is visible in the RoI distributions, consistent with the line plots, which showed localized underprediction of velocity in some wake regions. The $I_u$ histograms show that the hit rate for the full domain is consistently higher than for the RoI. This is because the full domain includes large areas of uniform turbulent flow outside the canopy, which are easily predicted and thus improve the overall metric.

    In summary, the model shows good predictive performance for both \( U_{\text{mag}} \) and \( I_u \) at \( 45^\circ \), with hit rates consistently above 88\% across all heights. The \( 0^\circ \) case shows slightly higher overall accuracy, particularly in the full flow field. A general trend of decreasing accuracy with height is observed for both wind directions and QoIs, likely due to the increasing complexity of three-dimensional flow structures caused by the buildings at lower heights. Notably, across the entire test set, the predictions for the \( 0^\circ \) layouts achieve the highest accuracy, while the \( 45^\circ \) layouts show the lowest. This discrepancy may be because the \( 0^\circ \) and \( 90^\circ \) configurations have similar geometric characteristics relative to the inflow, effectively doubling the number of training samples with these features and thus improving model capacity and accuracy for these wind directions. Including additional heights and more wind directions in the training data could help mitigate current limitations by allowing the model to better capture vertical and directional flow structures. Overall, the histograms and hit rates confirm the good performance of the model at these two wind directions. To asses the overall performance of the model, we quantify the errors by taking into account the whole test set in the next section.
    
\subsection{Comparison across all wind directions}

    The test set consists of 25 different configurations across 7 wind directions. Each of these 25 cases is augmented by flipping the corresponding images vertically along the $y$-axis, resulting in a total of 50 test cases. This augmentation doubles the test set size while preserving the physical validity of the flow fields, as symmetry about the $y$-axis is a reasonable assumption for our setup.

    To evaluate the performance of the U-net model, we use the following metrics: the hit rate, the NRMSE, and the MRE. All metrics are computed for the two QoIs. The results are presented in Table \ref{tab:performance_metrics_flow_pred}. For each wind direction case, the metrics are averaged across the three heights. In addition, the last row shows the overall average performance across all 50 test cases.

    \begin{table*}[htpb]
    \centering
    \caption{Performance metrics for different cases (HR = Hitrate)}
    \label{tab:performance_metrics_flow_pred}
    \footnotesize
    \begin{tabular}{l*{12}{c}}
    \toprule
    Cases & \multicolumn{2}{c}{\textbf{HR $U_{mag}$}} & \multicolumn{2}{c}{\textbf{HR $I_u$}} & \multicolumn{2}{c}{\textbf{NRMSE $U_{mag}$}} & \multicolumn{2}{c}{\textbf{NRMSE $I_u$}} & \multicolumn{2}{c}{\textbf{MRE $U_{mag}$}} & \multicolumn{2}{c}{\textbf{MRE $I_u$}} \\
    \cmidrule(lr){2-3} \cmidrule(lr){4-5} \cmidrule(lr){6-7} \cmidrule(lr){8-9} \cmidrule(lr){10-11} \cmidrule(lr){12-13}
    & Full & RoI & Full & RoI & Full & RoI & Full & RoI & Full & RoI & Full & RoI \\
    \midrule
    0 deg   & 97.3\% & 94.0\% & 95.3\% & 92.6\% & 0.033 & 0.050 & 0.030 & 0.036 & 7.3\% & 11.3\% & 5.3\% & 6.0\% \\
    45 deg  & 94.6\% & 96.3\% & 92.6\% & 89.6\% & 0.046 & 0.043 & 0.033 & 0.036 & 9.3\% & 11.3\% & 5.0\% & 7.0\% \\
    \midrule
    Overall & 95.1\% & 94.6\% & 93.6\% & 90.8\% & 0.044 & 0.048 & 0.033 & 0.035 & 9.1\% & 11.1\% & 5.2\% & 6.7\% \\
    \bottomrule
    \end{tabular}
    \end{table*}  

    For the full test set, the average hit rates for $U_{mag}$ and $I_u$ are 95.1\% and 93.6\% for the full flow field, and 94.6\% and 90.8\% for the RoI. These high hit rates show that the model predictions are consistently close to the ground truth, with the vast majority of points falling within the acceptable error bounds defined for each QoI. This is especially important given the diversity of the urban areas and wind directions, confirming the generalizability of the model.

    In terms of absolute error, the NRMSE values for the full flow field are 0.044 for $U_{mag}$ and 0.033 for $I_u$, while for the RoI, they increase slightly to 0.048 and 0.035, respectively. This trend is also reflected in the MRE values, which are 9.1\% and 5.2\% for $U_{mag}$ and $I_{u}$ in the full domain, and 11.1\% and 6.7\% in the RoI. The slightly higher errors in the RoI are expected, as these regions typically contain more complex flow features, such as wake interactions, flow separation, and shear layers around buildings. In contrast, outside of the urban area the flow is relatively uniform, which makes it easier to predict, and can decrease the errors. 

    Despite the increased complexity in the RoI, the U-net model shows robust performance. The hit rate remains close to or above 90\% for both QoIs in that region, indicating that the model remains accurate even inside the RoI. This is important, as studies of wind comfort, ventilation, and pollutant dispersion typically focus on regions within the urban canopy, where accurate flow predictions are crucial.

    Overall, the results validate the effectiveness of the proposed U-net architecture in predicting both velocity magnitude and turbulence intensity across a wide range of urban configurations and wind directions. The model shows high accuracy not only in the relatively uniform outer regions but also within the urban canopy, where the flow is influenced by the buildings. These findings suggest that the model has learned to generalize flow features across a diverse set of configurations.

\subsection{Limitations and future research avenues}

Although the proposed U-net model shows high predictive accuracy across various synthetic urban configurations and wind directions, several limitations should be acknowledged, along with promising directions for future research.
    
    First, the urban geometries used for training and evaluation are synthetic and relatively simple compared to real cities. Although this approach includes controlled diversity, it limits the capabilities of the model to be used in realistic urban environments. Future work should include high-resolution urban datasets with a higher level of detail derived from actual city layouts to assess generalizability.
    
    Second, the model uses 2D input slices to predict flow fields at multiple heights, based on only three vertical planes. While this approach is computationally efficient, it limits the ability of the model to fully capture three-dimensional flow dynamics, particularly at higher heights where vertical shear and wake interactions become more significant. Increasing the vertical resolution and including 3D spatial features, such as elevation maps, could improve the model and overcome this limitation.
    
    Third, the dataset does not account for several environmental factors that could significantly influence urban wind flow. In particular, variations in the atmospheric boundary layer, thermal stratification, and vegetation effects are not represented. The model has only been trained under neutral atmospheric boundary layer conditions. These simplifications limit its applicability to real scenarios where buoyancy effects, temperature gradients, and vegetative drag play important roles. Including these physical effects in the dataset or through physics-informed modeling techniques represents a direction for future research.
    
    Addressing these limitations will be essential for extending the applicability of the proposed model from controlled synthetic setups to more realistic and complex urban environments. This will be particularly important for applications in urban planning, pedestrian wind comfort analysis, and ventilation assessments.

\section{Conclusions}
    This paper introduces a Deep Neural Network (DNN) model to predict velocity magnitude and turbulence intensities in urban canopies at various heights. The urban canopies are scaled down by a factor of 1:100 to represent cities spanning 600 meters in both streamwise and spanwise directions, with building heights ranging between 10 meters and 50 meters in full scale. The model training was based on Large Eddy Simulations (LES) that include 252 different city configurations across 7 wind directions, ranging from $0^{\circ}$ to $90^{\circ}$ in $15^{\circ}$ increments. The dataset was divided into training, development, and test sets in an 80:10:10 ratio, which later was augmented by flipping the images in the spanwise direction. This results in train:dev:test ratio of 404:50:50. An evaluation using the DNN model takes approximately $\mathcal{O}1$ sec on a single GPU, compared to LES, which needs $\mathcal{O}10$ hours on 32 CPUs. The fast and accurate predictions make the model suitable for 
    urban planning, pedestrian wind comfort analysis, and ventilation assessments, particularly when new buildings are planned.
    
    The DNN model employs a U-net architecture, consisting of four convolutional blocks for downsampling and four for upsampling. Each block includes two convolutional layers with ReLU activation and batch normalization. A Spatial Attention Module is used in the skip connections to enhance encoder features before decoding. The input consists of a binary representation of the city, the Signed Distance Function, and its gradient at three heights, forming a $256\times256\times9$ tensor. The output is a $256\times256\times6$ tensor containing predictions of velocity magnitude and turbulence intensity at three heights each.
    
    The loss function included three components: 1) the root-mean-square error (RMSE) between the predicted and estimated Quantities of Interest (QoIs), 2) the RMSE of the gradient magnitude based on the output images, and 3) a regularization term using the $L_2$ norm applied to the U-net model weights. The hyperparameters tuned during the training included the number of reference channels in each convolutional block, kernel size, weight for the gradient magnitude term, regularization weight, batch size, and learning rate. The optimal model was selected based on performance on the dev set, and its predictive capabilities were evaluated on the test set.
    
    The evaluation of the predictive capabilities of the DNN model was conducted using the test set, which included 50 cases. The results showed that the model accurately predicted the velocity magnitude and turbulence intensity within urban canopies, with an overall mean relative error (MRE) of 9.3\% for velocity magnitude and 5.2\% for turbulence intensity, when the full flow field is considered. When a region of interest (RoI) at the center of the city if taken into account, the MRE is 11.1\% and 6.7\%, respectively. This highlights that the model can accurately predict the complex flow patterns within the urban canopy, where wakes and shear layers are present. 
    
    The model achieved highest accuracy for inflow directions aligned with the building orientation ($0^\circ$ and $90^\circ$), with slightly reduced performance at $45^\circ$, likely due to the more limited representation of diagonal wind cases in the training data. This suggests that increasing the number and diversity of training samples, especially for intermediate wind angles, can further improve model generalization.
    
    Despite its promising performance, several limitations remain. The model was trained and validated on synthetic urban configurations, which, although diverse, lack the geometric complexity of real cities. The 2D input representation and use of only three vertical planes limit the ability to capture fully three-dimensional flow features, especially at higher heights. Additionally, the dataset includes neutral atmospheric boundary layer conditions and does not account for environmental factors such as atmospheric boundary layer variability, thermal stratification, or vegetation effects. 

    Future research will address these issues by extending the model to more realistic urban setups, including non-rectilinear buildings, variable roof structures, terrain elevation, and vegetation. Enhancing vertical resolution and incorporating 3D spatial information, such as elevation maps, will be explored to improve predictions. In addition, integrating environmental conditions such as thermal effects and atmospheric stability, and embedding physical constraints into the learning process, may improve generalizability and the robustness of the predictions. These improvements will be key in deploying deep learning models for real-world urban flow applications.

\section*{CRediT authorship contribution statement}
 \textbf{Themistoklis Vargiemezis}: Data curation, Formal analysis, Methodology, Investigation, Validation, Visualization, Writing – original draft, Writing – review \& editing. \textbf{Catherine Gorl\'e}: Conceptualization, Formal analysis, Funding acquisition, Methodology, Supervision, Writing – review \& editing.

\section*{Declaration of competing interest}
The authors declare that they have no known competing financial interests or personal relationships that could have appeared to influence the work reported in this paper.

\section*{Acknowledgements}
This material is based upon work supported by the National Science Foundation under Grant Number 1749610 and used the Extreme Science and Engineering Discovery Environment (XSEDE), which is supported by National Science Foundation grant number CI-1548562.



 \bibliographystyle{elsarticle-harv} 
 \bibliography{main}





\end{document}